\documentclass[letterpaper, 10 pt]{IEEEtran}%
\usepackage{amsmath,amsfonts}

\usepackage[english]{babel}
\usepackage{array}
\usepackage[caption=false,font=normalsize,labelfont=sf,textfont=sf]{subfig}
\usepackage{textcomp}
\usepackage{stfloats}
\usepackage{url}
\usepackage{verbatim}
\usepackage{graphicx}
\usepackage{cite}
\usepackage{units}
\usepackage{dsfont}
\usepackage{booktabs}
\usepackage[dvipsnames]{xcolor}
\usepackage{xfrac}
\usepackage{epstopdf}
\usepackage{algorithm}
\usepackage{algpseudocode}
\usepackage{circuitikz}
\usepackage{tikz}
\usepackage{tikz-uml}

\usetikzlibrary{shapes, arrows,calc,positioning }
\usepackage{eurosym}
\usepackage{adjustbox}
\usepackage{multirow}
\usepackage{xcolor}

\definecolor{dataColor}{RGB}{0, 102, 204}
\definecolor{simColor}{RGB}{255, 94, 77}
\definecolor{infraColor}{RGB}{96, 143, 159}
\definecolor{opColor}{RGB}{34, 139, 34}

\usepackage{amsthm}

\newtheorem{theorem}{Theorem}[section]

\usepackage{color, colortbl}

\definecolor{expRR}{HTML}{7498C8}
\definecolor{expRO}{HTML}{FECB91}
\definecolor{expOR}{HTML}{A6D9A6}
\definecolor{expOO}{HTML}{F4766C}

\newcommand{\mr}[1]{\mathrm{#1}}
\newcommand{\rmax}{\mathrm{max}}

\title{Simultaneous Optimization of Electric Ferry Operations and Charging Infrastructure}

\author{Juan Pablo Bertucci$^{1}$, Theo Hofman$^{1}$, Mauro Salazar$^{1}$%
\thanks{$^{1}$Department of Mechanical Engineering, Control System Technology,
        Eindhoven University of Technology, 5600 MB  Eindhoven, The Netherlands
        {\tt\small \{j.p.bertucci\}@tue.nl}}%
}

\begin{document}

\maketitle
\thispagestyle{empty}
\pagestyle{empty}

\begin{abstract}
Electrification of marine transport is a promising solution to reduce sector greenhouse gas emissions and operational costs.
However, the large upfront cost of electric vessels and the required charging infrastructure can be a barrier to the development of this technology.
Optimization algorithms that jointly design the charging infrastructure and the operation of electric vessels can help to reduce these costs and make these projects viable.
In this paper, we present a mixed-integer linear programming optimization framework that jointly schedules ferry operations, charging infrastructure and ship battery size.
We analyze our algorithms with the case of the China Zorrilla, the largest electric ferry in the world, which will operate between Buenos Aires and Colonia del Sacramento in 2025.
We find that the joint system and operations design can reduce the total costs by 7.8\% compared to a scenario with fixed power limits and no port energy management system.
\end{abstract}

\section{Introduction}\label{sec:introduction}
Maritime electrification is gaining traction due to its potential to reduce greenhouse gas emissions and lower operational costs~\cite{DNVGL2020,SiemensBellona2019}. Recent deployments of battery-electric ferries on short- to medium-range routes have demonstrated their technical and economic viability. Notable examples include Denmark's \emph{Ellen}~\cite{EllenProjectReport2019} and Norway's \emph{MF Ampere}~\cite{AmpereCaseStudy2018}, where high-capacity battery systems and rapid-charging infrastructure enabled fully electric, zero-emission operations. Similar benefits were observed in retrofitting existing vessels, such as the Swedish-Danish ferries \emph{Tycho Brahe} and \emph{Aurora}~\cite{ForSeaConversion2021}, which achieved significant reductions in operating costs. Additionally, the upcoming deployment of the largest fully electric ferry \emph{China Zorrilla} in the R\'{\i}o de la Plata in late 2025 further supports the trend toward maritime electrification.
However, broader implementation of electric ferries is limited by the high upfront investments and substantial power infrastructure requirements, as well as the technichal limitations of range. Traditionally, infrastructure sizing and ferry operations are optimized separately, neglecting their interdependencies. This fragmented approach can lead to oversized infrastructure, operational inefficiencies, and higher costs.
\subsubsection*{Related literature}
This work relates to two main research streams: optimal all-electric ferry operations and optimized charging infrastructures.
The first stream focuses on energy management and power dispatch on the vessel side. Genetic algorithms have been used to balance economic and environmental scheduling~\cite{shangEconomicEnvironmentalGeneration2016}, while nonlinear optimization formulations have been proposed for comparative analysis of fuel-cell powered vessels~\cite{banaeiComparativeAnalysisOptimal2020}. Other studies examine the influence of battery capacity on hybrid propulsion~\cite{choiEnergyManagementStrategies2024}, multi-energy ship power systems~\cite{fangOptimizationBasedEnergyManagement2021}, and propose coordinated route planning and energy dispatch~\cite{heinCoordinatedOptimalVoyage2021}. Simulation based economic analyses were developed in ~\cite{kistnerPotentialsLimitationsBatteryelectric2024a} finding the route length limits full electrification significantly. Methods that simultaneously size components and schedule energy flows in~\cite{letafatSimultaneousEnergyManagement2020} show savings in operational costs, while case studies validate cost competitiveness of electrification under realistic load profiles~\cite{royOperationalCostOptimization2022,pesaElectricFerryFleet2025}.
The second stream addresses charging logistics and port-side infrastructure. Research has examined cold ironing and onshore battery systems to navigate grid limitations and reduce emissions for short-berthing ships~\cite{abubakarTwostageEnergyManagement2024a}, with additional work detailing system designs and performance for high-power demands~\cite{capraraColdIroningBattery2022}. Life-cycle cost models inform optimal sizing of local renewable generation and storage~\cite{colarossiOptimalSizingPhotovoltaic2023}. Using storage for grid support in trasnportation charging infrastructure applications has been recognized as key to decrease costs in other transportation domains~\cite{EMSBertucci2025}.
Despite these advancements, the literature still lacks a comprehensive methodology that addresses large-capacity, high-speed ferry operations simultaneously accounting for long-term battery degradation, and shore-side charging infrastructure design with global optimality guarantees.
\begin{figure}[!t]
\centering
\includegraphics[width=\columnwidth]{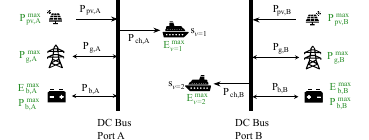}
\caption{Problem sketch showing the main optimization design variables in green. Maximum photovoltaic production $P_{\mathrm{pv},i}$, peak grid connection $P_{\mathrm{g},i}$ and battery storage capacity $E_{\mathrm{b},i}$ and power $P_{\mathrm{b},i}$. On the vessel side, the maximum battery $E^{\mathrm{max}}_{v}$. Operational variables include the power dispatch variables from each source $P_{\mathrm{b},i}$,$P_{\mathrm{g},i}$,$P_{\mathrm{pv},i}$, and the operation speed of each vessel $s_{v}$.}
\end{figure}
\subsubsection*{Statement of Contributions}
To address this gap, this paper introduces a Mixed-Integer Linear Programming (MILP) optimization framework that simultaneously determines optimal ferry scheduling, battery sizing, and port-side charging infrastructure design. The proposed method guarantees globally optimal solutions and explicitly integrates operational constraints with infrastructure investment decisions.
\subsubsection*{Organization}
The remainder of this paper is structured as follows: Section~\ref{sec:meth} presents the MILP problem formulation, including the battery, charging, and scheduling models. Section~\ref{sec:res} reports numerical results for different design scenarios of our case study. Finally, Section~\ref{sec:dis} concludes the paper and highlights directions for future research.

\section{Methodology}\label{sec:meth}
This section develops the optimization formulation, we describe the sets, constraints, and decision variables used in the optimization problem.
\subsection{Definitions}
We consider a set of vessels $v\in\mathcal{V}$ providing service between ports $i \in \mathcal{I}$. Each pair of locations $(i,j)$ with $i,j \in \mathcal{I}$ has distance $d_{ij}$.
Each vessel $v$ is characterized by a battery size $E^{\rmax}_{v}$ (MWh), weight $w_{v}$ (ton), and a lumped friction constant $k_v$.
The planning horizon is discretized as $\mathcal{T} = \{1,\dots,T\}$, with each period having duration $\tau$.
An itinerary $\mathcal{L}^{v}$ for vessel $v$ is composed of legs $\ell \in \mathcal{L}^{v}$. Each leg specifies an origin $i_{\ell}$, destination $j_{\ell}$, distance $d_{\ell}$, vessel displacement $w_{\ell}$, and time windows for arrival and departure $\bigl(t_{v,\ell}^{\mr{dep,ear}},\,t_{v,\ell}^{\mr{dep,lat}}\bigr)$ and $\bigl(t_{v,\ell}^{\mr{arr,ear}},\,t_{v,\ell}^{\mr{arr,lat}}\bigr)$. The tuple $(v,l)$ therefore defines univocally origin, destination, and times of each vessel.
\subsection{Decision Variables}
Each site $i\in\mathcal{I}$ includes a local energy system with a peak grid connection $P^{\max}_{\mr{g},i}$, maximum installed photovoltaic (PV) generation $P^{\rmax}_{\mr{pv},i}$, and on-site storage $E^{\rmax}_{\mr{b},i}$ which are optimized in this formulation.
We let $E_{\mr{b},i}(t)$ $(\text{MWh})$ be the stationary storage state of energy, and $P_{\mr{b},i}(t)$ $(\text{MW})$ the net power at a location $i$ at time $t$, which we split into energy received $P_{\mr{b},i}^{+}(t)$ and dispatched ,$P_{\mr{b},i}^{-}(t)$ to keep the formulation linear.
Similarly, the site can buy or sell power to/from the grid, $P_{\mr{g},i}^{+}(t)$,$P_{\mr{g},i}^{-}(t)$, with net grid flow given by $P_{\mr{g},i}(t)= P_{\mr{g},i}^{+}(t)$-$P_{\mr{g},i}^{-}(t)$.
Regarding the PV production, we define $P_{\mr{pv},i}(t)$ as the power produced by the PV system at site $i$ at time $t$.
Each vessel $v$ has a battery size $E_{v}^{\mr{max}}$ (\unit{MWh}) and $E_{v}(t)$ $(\text{MWh})$ be the vessel's state of energy at time $t$ and $P^{\mr{ch}}_{v,i}(t)$ the charging power from site $i$ to vessel $v$ at time $t$.
The travel time $t_{v,\ell}^{\mr{travel}}$ and speed $s_{v,\ell}$ can be varied by each vessel.
\subsection{Constraints}
We first develop the vessel operational constraints, followed by the shore-side contraints at each port.
\subsubsection{Vessel Operation Constraints}
Each leg $\ell$ of vessel $v$ has earliest and latest departure times $\bigl(t_{\ell}^{\mr{dep,ear}},\,t_{\ell}^{\mr{dep,lat}}\bigr)$ and arrival times $\bigl(t_{\ell}^{\mr{arr,ear}},\,t_{\ell}^{\mr{arr,lat}}\bigr)$.
The actual departure and arrival times satisfy
\begin{equation}
    \label{eq:firsteq}
    t_{v,\ell}^{\mr{dep,ear}}
    \;\;\le\;\;
    t_{v,\ell}^{\mr{dep,act}}
    \;\;\le\;\;
    t_{v,\ell}^{\mr{dep,lat}}
    \quad
    \forall\,v,\ell,
    \end{equation}
\begin{equation}
    t_{v,\ell}^{\mr{arr,act}}
    \;=\;
    t_{v,\ell}^{\mr{dep,act}}
    \;+\;
    t_{v,\ell}^{\mr{travel}},
    \quad
    \forall\,v,\ell.
    \end{equation}
Each travel time $t_{v,\ell}^{\mr{travel}}$ is determined by the vessel speed $s_{v,\ell}$, and distance $d_{v,\ell}$ for each trip leg. We enforce this with the following constraints:
\begin{align}
    s_{v,\ell}\;t_{v,\ell}^{\mr{travel}} \;=\; d_{v,\ell},
    \quad
    s_{v,\ell}^{\mr{min}}\;\le\;s_{v,\ell}\;\le\;s_{v,\ell}^{\mr{max}},
    \quad
    \forall\,v,\ell.
    \end{align}
from this, we can determine the consumption per leg $E_{v,\ell}^{\mr{cons}}$ derived from the hydrodynamic resistance \cite{larssonShipResistanceFlow2010}:
\begin{equation}
    E_{v,\ell}^{\mr{cons}}
    \;=\;
    k\;d_{v,\ell}\;s_{v,\ell}^{2}\;w_{v,\ell}^{\tfrac{2}{3}},
    \quad
    \forall\,v,\ell.
    \end{equation}
To keep our formulation linear we formulate a convex envelope that slightly overestimates consumption, and is thus on the conservative side.
Then, for each leg:
\begin{equation}
    E_{v,\ell}^{\mr{arr}}
    \;\;\ge\;\;
    E_{v,\ell}^{\mr{dep}}
    \;+\;
    E_{v,\ell}^{\mr{char}}
    \;-\;
    E_{v,\ell}^{\mr{cons}},
    \quad
    \forall\,v,\ell,
    \end{equation}
And ensure the vessel's state of charge at the end of each leg is equal to the state of charge at the beginning of the next leg:
\begin{equation}
    E_{v,\ell}^{\mr{dep}} \;=\; E_{v,\ell-1}^{\mr{arr}},
    \quad
    \forall\,v,\;\ell>1,
    \end{equation}
And link the energy charged prior to each leg to the charging decisions $P_{v,\ell}^{t}$:
\begin{equation}
    E_{v,\ell}^{\mr{char}}
    \;=\;
    \,\tau \, \sum_{t \,\in\, T_{\ell}^{v}}
    P_{v,\ell}^{t},
    \quad
    \forall\,v,\ell.
    \end{equation}
We also enforce depth-of-discharge and maximum energy constraints
\begin{equation}
    E_{v,\ell}^{\mr{arr}}
    \;\ge\;
    E_{v}^{\mr{min}},
    \quad
    E_{v,\ell}^{\mr{dep}}
    \;+\;
    E_{v,\ell}^{\mr{char}}
    \;\le\;
    E_{v}^{\mr{max}},
    \quad
    \forall\,v,\ell.
    \end{equation}
We also enforce periodic constraints at the start and end of the trip:
\begin{align}
    E_{v,\,1}^{\mr{arr}} &\;\;\le\;\;\phi_v\,E_{v}^{\mr{max}}, \\
    E_{v,\,|\mathcal{L}_v|}^{\mr{arr}} &\;\;\ge\;\;\phi_v\,E_{v}^{\mr{max}},
    \quad
    \forall\,v.
    \end{align}
where $\phi_v$ is a parameter that defines the maximum state of charge at the beginning of the first leg and the minimum state of charge at the end of the last leg as a percentage of the battery capacity.
To enforce charging only during the time periods the vessel is moored, we define
\begin{equation}
    0
    \;\;\le\;\;
    P^{\mr{ch}}_{v,\ell}(t)
    \;\;\le\;\;
    M\,Z_{v,\ell}(t),
    \quad
    \forall\,v,\ell,\;t,
    \end{equation}
which bounds the aggregated charging power by a large constant ($M$) whenever the binary variable $Z_{v,\ell}(t)=1$ indicates that charging is allowed; otherwise the upper limit is forced to zero. The charging power $P^{\mr{ch}}_{v,l}(t)$ is defined as the sum of the power from all sources to the vessel $v$ at leg $l$ and time $t$:
\begin{align} \nonumber
\label{eq:power-charging}
P^{\mr{ch}}_{v,\ell}(t)
\;=\;
\sum_{i\in\mathcal{I}}
\Bigl[
P^{\mr{g2v}}_{i,v,\ell}(t)
\;+\;
P^{\mr{pv2v}}_{i,v,\ell}(t)
\;+\;
P^{\mr{b2v}}_{i,v,\ell}(t)
\Bigr]
\;\delta_{i,v,\ell}(t), \\
\quad
\forall\,v,\ell,\;t,
\end{align}
where $P^{\mr{g2v}}_{i,v,l}(t)$ corresponds to the power from the grid to the vessel, $P^{\mr{pv2v}}_{i,v,l}(t)$ is the power from the PV to the vessel, and $P^{\mr{b2v}}_{i,v,l}(t)$ is the power from the battery to the vessel, and $\delta_{i,v,l}(t)$ is an indicator variable with value 1 when the vessel $v$ is at site $i$ during leg $l$ at time $t$, and zero otherwise.
We link these power variables to the power drawn from each port's energy system in the equations~\eqref{eq:power-charging-port}.
We then define the net energy charged for a given vehicle/leg combination:
\begin{equation}
    E_{v,\ell}^{\mr{char}}
    \;=\;
    \tau \sum_{t} P^{\mr{ch}}_{v,\ell}(t),
    \quad
    \forall\,v,\ell.
    \end{equation}
imposing consistency between the decision variable $E_{v,\ell}^{\mr{char}}$ and the net time-integral of power.
Finally, to ensure that charging times correspond with the operational departure and arrival times, we write:
\begin{equation}
t \;\;\le\;\;
t_{v,\ell}^{\mr{dep,act}}
\;+\;
M\bigl[1 - Z_{v,\ell}(t)\bigr],
\quad
\forall\,v,\ell,\;t,
\end{equation}
\begin{equation}
    t \;\;\ge\;\;
t_{v,\ell-1}^{\mr{arr,act}}
\;-\;
M\bigl[1 - Z_{v,\ell}(t)\bigr],
\quad
\forall\,v,\ell,\;t.
\end{equation}
where $M$ is a suitably large constant. These inequalities ensure that $t$ falls into the interval where charging is feasible according to operational schedules.
\subsubsection{Port (Shore-Side) Constraints}
The delivered power at each location cannot exceed the maximum installed charging power $P_{\mr{g},i}^{\mr{max}}$
\begin{equation}
P_{v,i}(t)\;\le\;P_{\mr{ch},i}^{\mr{max}},
\quad
\forall\,v,\;i,\;t.
\end{equation}
We link the power drawn from the grid at each time $t$ to the power supplied to the vessels through:
\begin{align}
\label{eq:power-discharging-port}
P_{\mr{g},i}^{+}(t)
\;=\;
\sum_{v\in\mathcal{V},\ell\in\mathcal{L}^{v}}
\Bigl[
P^{\mr{g2v}}_{i,v,\ell}(t)
\Bigr]
\;+\;
P^{\mr{g2b}}_{i}(t)
\quad
\forall\,i,t,
\end{align}
and the power supplied to the grid through:
\begin{align}
\label{eq:power-charging-port}
P_{\mr{g},i}^{-}(t)
\;=\;
P^{\mr{pv2g}}_{i}(t)
\;+\;
P^{\mr{b2g}}_{i}(t)
\quad
\forall\,i,t,
\end{align}
The available solar power per installed kW of PV at site $i$ in period $t$ is denoted by $\gamma_{i}(t)$.
Thus, PV generation is constrainted by
\begin{equation}
    P_{i}^{\mr{pv}}(t)
    \;\le\;
    P^\mr{pv,max}_{i}\;\gamma_{i}(t),
    \quad
    \forall\,i,\;t.
    \end{equation}
We link the generation to the different uses of the power generated by the PV system at each site $i$ through:
\begin{equation}
    \begin{split}
    \label{eq:power-charging-port2}
    P_{\mr{pv},i}(t)
    \;=\;
    \sum_{v\in\mathcal{V},\ell\in\mathcal{L}^{v}}
    \Bigl[
    P^{\mr{pv2v}}_{i,v,\ell}(t)
    \Bigr]
    \;+\; \\
    P^{\mr{pv2b}}_{i}(t)
    \;+\;
    P^{\mr{pv2g}}_{i}(t)
    \quad
    \forall\,i,t,
\end{split}
\end{equation}
We limit the design variables for the installed PV capacity $P_{\mr{pv,i}}^{\mr{max}}$ to be between zero and the maximum installed capacity $P^{\mr{M}}_{\mr{pv}}$:
\begin{equation}
0
\;\le\;
P_{\mr{pv},i}^{\mr{max}}
\;\le\;
P^{{\mr{M}}}_{\mr{pv}},
\quad
\forall\,i,
\end{equation}
and the instantaneous power is then limited via:
\begin{equation}
P_{\mr{pv},i}(t)
\;\le\;
P_{\mr{pv},i}^{\mr{max}}
\;\gamma_{i}(t),
\quad
\forall\,i,\;t.
\end{equation}
For the power drawn from the stationary storage:
\begin{align}
\label{eq:power-charging-port3}
P_{\mr{b},i}^{+}(t)
\;=\;
\sum_{v\in\mathcal{V},\ell\in\mathcal{L}^{v}}
\Bigl[
P^{\mr{b2v}}_{i,v,\ell}(t)
\Bigr]
\;+\;
P^{\mr{b2g}}_{i}(t)
\quad
\forall\,i,t,
\end{align}
And the power supplied to the storage:
\begin{align}
\label{eq:power-charging-port4}
P_{\mr{b},i}^{-}(t)
\;=\;
P^{\mr{g2b}}_{i}(t)
\;+\;
P^{\mr{pv2b}}_{i}(t)
\quad
\forall\,i,t,
\end{align}
We then write for each site $i$, the stationary battery state of charge $E_{\mr{b},i}(t)$ dynamics as:
\begin{equation}
    \begin{split}
    E_{\mr{b},i}(t+1)
\;=\;
E_{\mr{b},i}(t)
\;+\;
\tau\,\Bigl[\eta_{\mr{c,b}}\,P_{\mr{b},i}^{+}(t)\\
\;-\;
\tfrac{1}{\eta_{\mr{d,b}}}\,P_{\mr{b},i}^{-}(t)\Bigr],
    \end{split}
\end{equation}
We bound the operation of the battery by the maximum and minimum state of charge :
\begin{equation}
    E_{\mr{b},i}^{\mr{min}}
    \;\;\le\;\;
    E_{\mr{b},i}(t)
    \;\;\le\;\;
    E_{\mr{b},i}^{\mr{max}},
    \quad
    \forall\,i,\;t,
    \end{equation}
and the maximum power limits for safe charging and discharging:
    \begin{equation}
    -P_{i}^{\mr{max}}
    \;\le\;
    P_{\mr{b},i}(t)
    \;\le\;
    P_{i}^{\mr{max}},
    \quad
    \forall\,i,\;t.
    \end{equation}
\subsection{Costs and Objective Function}
The unit costs for installing stationary energy storage are given by $C_{\mr{b,i}}$ (k\$ per MWh).
The cost of installing PV is $C_{\mr{pv}}$ (k\$ per MW), and the grid connection cost is $C_{\mr{g}}$ (k\$ per kW).
The electricity purchase price at site $i$ and time $t$ is $p_{i}(t)$ (k\$ per kWh), while $\lambda_{i}\,p_{i}(t)$ is the revenue obtained per kWh injected back to the grid.
The unit costs vessel storage are given by $C_{\mr{b,v}}$ (k\$ per MWh).
We minimize the total costs, composed of capital and operational expenses.
The capital expenses consider the costs of the energy storage, grid connection, and PV generation for each site:

\begin{equation}
C_{cap,i} =
C_{\mr{b,i}}\,E_{\mr{b},i}^{\mr{max}} +
C_{\mr{pv}}\,P_{\mr{pv},i}^{\mr{max}} +
C_{\mr{g}}\,P_{\mr{g},i}
\end{equation}
and the cost for the batteries on board the vessels:
\begin{equation}
    C_{cap,v} = C_{\mr{b,v}}\,E_{v}^{\mr{max}}
\end{equation}
and the operating cost for each site
\begin{equation}
    \label{eq:cost-operating}
    C_{\mr{op},i} =
\tau
\sum_{t=1}^{T}
p_{i}(t) \, \big[\,P_{\mr{g},i}^{+}(t)
-\lambda_{i} \,P_{\mr{g}i}^{-}(t)\big].
\end{equation}
Thus, we define the logistic operator's optimization problem to be centered around minimizing the total costs, so we define our objective function as
\begin{equation}
    \begin{split}
    J(E_{\mr{b},i}^{\mr{max}},P_{\mr{b},i}^{\mr{max}},P_{\mr{pv},i}^{\mr{max}},P_{\mr{g},i},E_{v}^{\mr{max}},P_{i}) \\
    =\sum_{v} C_{\mr{cap},v} + \sum_{i\in\mathcal{I}} C_{\mr{cap},i} + \sum_{i\in\mathcal{I}} C_{\mr{op},i}
\end{split}
\end{equation}
\subsection{Problem Formulation}
We can thus write the minimization of the operator's cost as follows:

\textbf{Problem 1: Joint Infrastructure \& Energy Management}
\textit{Given sets of sites $\mathcal{I}$, vessels $\mathcal{V}$ with each vessel $v$ having itineraries $\mathcal{L}_{v}$,
the energy storage ($E_{\mr{b},i}^{\mr{max}},P_{\mr{b},i}^{\mr{max}}$), grid connection ($P_{\mr{g},i}$), and PV capacity ($P_{\mr{pv,i}}^{\mr{max}}$), as well as the power dispatch $\bigl\{P_{\mr{b},i}(t),\,P_{\mr{pv},i}(t)\bigr\}$, state of charge of the battery storage $E_{\mr{b},i}$ and vessel charging schedules $P_{v,\ell}^{t}$ minimizing total cost ($J$) result from:}
\begin{equation}
\label{eq:final-prob}
\begin{aligned}
\min\;\; & J\bigl(E_{\mathrm{b},i}^{\mathrm{max}},\,
P_{\mathrm{b},i}^{\mathrm{max}},\,
P_{\mathrm{pv},i}^{\mathrm{max}},\,
P_{\mathrm{g},i}^{t},\,
P_{\mathrm{b},i}^{t},\,
P_{\mathrm{pv},i}^{t},\,
E_{\mathrm{b},i}^{t},\,
P_{v,\ell}^{t}\bigr) \\
\text{s.t.} \quad
& \eqref{eq:firsteq} - \eqref{eq:cost-operating}, \\
& E_{\mathrm{b},i}^{\mathrm{max}},\; P_{\mathrm{b},i}^{\mathrm{max}},\; P_{\mathrm{pv},i}^{\mathrm{max}}
\quad \forall i \in \mathcal{I}, \\
& P_{\mathrm{g},i}^{t},\; P_{\mathrm{b},i}^{t},\; P_{\mathrm{pv},i}^{t},\; E_{\mathrm{b},i}^{t}
\quad \forall i \in \mathcal{I},\, t \in \mathcal{T}, \\
& P_{v,\ell}^{t} \in \mathbb{R}_{+}
\quad \forall v \in \mathcal{V},\, \ell \in \mathcal{L}_v,\, t \in \mathcal{T}, \\
& Z_{v,\ell,t} \in \{0,1\}
\quad \forall v \in \mathcal{V},\, \ell \in \mathcal{L}_v,\, t \in \mathcal{T}.
\nonumber
\end{aligned}
\end{equation}
\noindent
Problem~1 is a MILP that can be solved globally to optimality by standard off-the-shelf optimization solvers. Considering the objective of minimizing cost, the convex relaxation in Problem~\eqref{eq:final-prob} is lossless (i.e., complementarity between positive and negative power components) for any $\lambda_{i}\in(0,1]$.
\subsection{Discussion}
A few comments are in order. First, the study assumes that vessel displacement varies only slightly with changes in battery configuration. This simplification aids in the modeling process but may warrant further investigation in cases where battery sizing significantly affects the vessel's weight distribution.
Second, the MILP scalability to a large fleet of vessels is limited. Future research should explore decomposition methods or alternative solution strategies that can handle larger fleet sizes more efficiently.
Finally, extending the framework to incorporate Vehicle-to-Grid (V2G) services is straightforward. However, such an extension was omitted from this paper due to the absence of detailed turnaround times and the lack of established incentive structures in the case study markets.

\section{Case Study}\label{sec:case}
\begin{figure}[!t]
\centering
\includegraphics[width=0.8\linewidth]{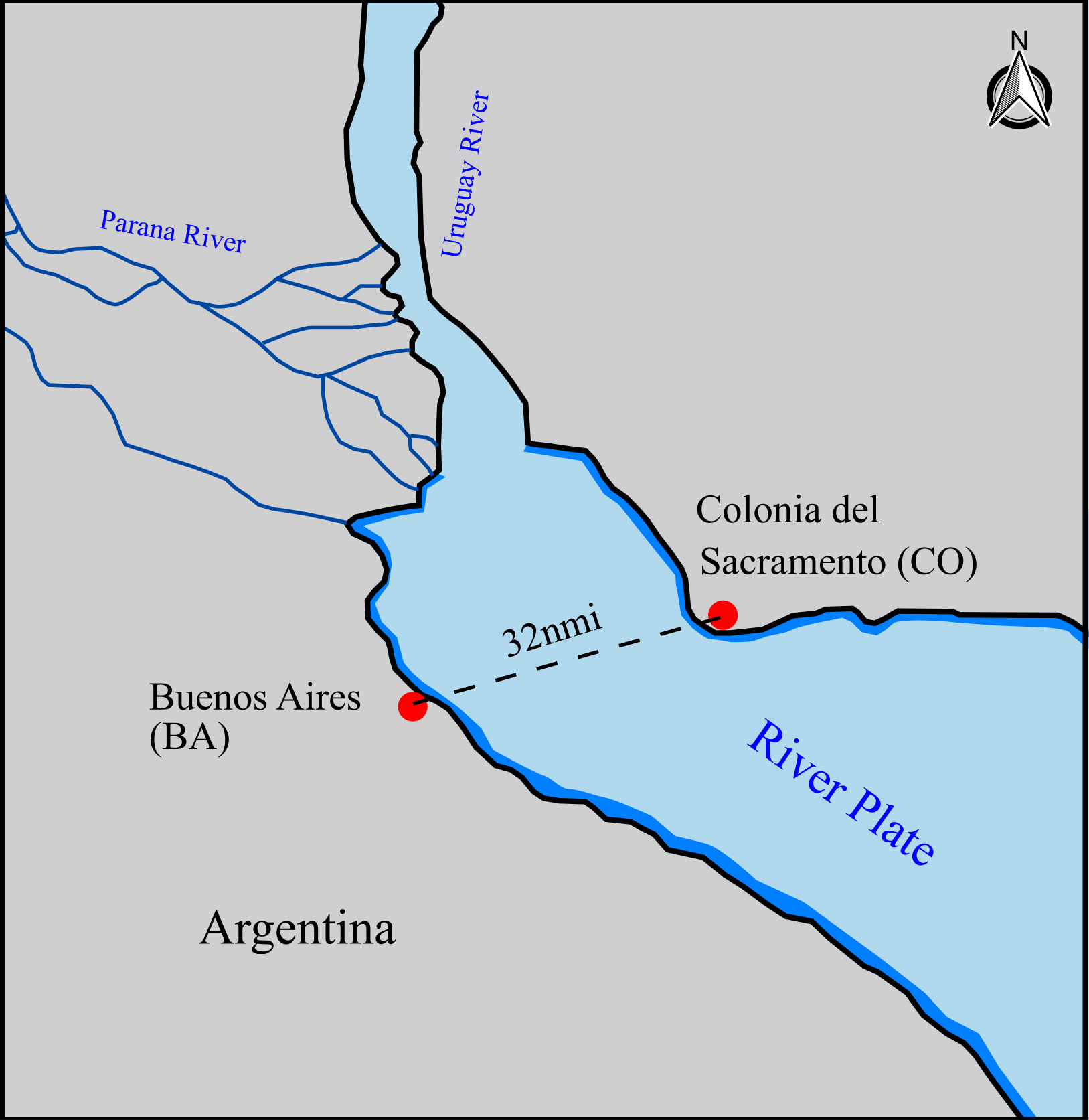}
\caption{The route to be covered by the China Zorrilla ferry between the port of Buenos Aires (BA) and port of Colonia del Sacramento (CO).}
\label{fig:route}
\end{figure}

In this section we describe the input parameters and experiments conducted to test the optimization algorithm. . We use the data for the CZ, assuming a future service of two ferries working to cover all the current demand between Colonia (CO) and Buenos Aires (BA), as illustrated in Fig.~\ref{fig:route}.
The vessel is planned to have capacity for 2100 passengers and 225 vehicles. The catamaran twin hulls are mostly aluminium, with a total length of 130\unit{m}, beam of 32\unit{m} and draft of 3.5\unit{m}.
The total battery capacity will be \unit[40]{MWh}, powered by 8 axis electric motors, optimized for speeds of 25\unit{kn}.
The charging infrastructure will have a DC shore charging system with a maximum connection of 15\unit{MW} following from communications with the service operator \cite{bqbMainPage}.
The service is planned to have two daily frequencies, with a total of 4 trips per day between both ports \footnote{Source: Buquebus ( https://buquebus.com/arribosypartidas)}.
Table~\ref{tab:parameters} summarizes the main parameter used in our analysis.
Historical values for solar power generation $P_\mathrm{PV}$ are obtained from \cite{PVGIS} and callibrated with local power generation data.
Energy prices were obtained for the simulated week and average between 64 and 135 per \unitfrac{USD}{MWh} for Buenos Aires\footnote{Source: CAMMESA (https://cammesaweb.cammesa.com)} and Colonia\footnote{Source: UTE (https://ute.com.uy/)} respectively.
\begin{table}[htbp]
  \centering
  \caption{Model parameters used in the case study.}
  \label{tab:parameters}
  \renewcommand{\arraystretch}{1.2}
  \begin{adjustbox}{max width=\linewidth}
    \begin{tabular}{lll|lll}
      \toprule
      \textbf{Var.} & \textbf{Value} & \textbf{Unit} & \textbf{Var.} & \textbf{Value} & \textbf{Unit} \\
      \midrule
      $\tau$             & 15      & min.            & $\lambda_i$              & 0.2       & min. \\
      $T_{\mathrm{start}}$         & 01.11.2023 & -- & $T_{\mathrm{end}}$                & 08.11.2023 & -- \\
      \midrule
      $E_{\mathrm{b},i}^{\mathrm{max}}$  & 50   & MWh    & $C_{\mathrm{b}}$    & 250      & USD/kWh \\
      $E_{\mathrm{v}}^{\mathrm{max}}$    & 50   & MWh    & $E_{\mathrm{v}}^{\mathrm{min}}$   & 15 & MWh \\
      \midrule
      $C_{\mathrm{veh}}^{\mathrm{batt}}$  & 400   & USD/kWh       & $\alpha_{\mathrm{resale}}$            & 0.2     & -- \\
      $C_{\mathrm{pv}}$    & 850  & USD/kW     & $P_{\mathrm{pv}}^{\mathrm{max}}$       & 5  & MW \\
      \midrule
      $p_{e}^{\mathrm{avg}}$     & [\,0.09 , 0.135\,] & USD/kWh   & $d_{\mathrm{BA,CO}}$        & 32      & nmi \\
      \midrule
      $v^{\mathrm{min}}$       & 10    & kn    & $v^{\mathrm{max}}$        & 25     & -- \\
      $t_{\mathrm{travel}}^{\mathrm{max}}$ & 2.50   & h   & $\mathrm{E}_{\mathrm{start}}$      & 0.80   & -- \\
      \midrule
      $\phi_{v}$          & 0.50   & --   & $\mathrm{E}_{\mathrm{stat}}^{\mathrm{start}}$ & 0.80 & -- \\
      $\phi_{\mr{b},i}$   & 0.80   & --   & $P_{i}^{\mr{max}}$       & 15 & MW \\
      \midrule
      $T^{\mr{amort}}_\mathrm{infra}$  & 15 & years    & $T^{\mr{amort}}_\mathrm{vessels}$ & 10 & years \\
      $s_v^{\mathrm{max}}$      & 30 & kn       & $s_v^{\mathrm{min}}$ & 10 & kn \\
      $t_{\mr{travel},v}^{\mathrm{max}}$ & 1.5 & hours & $t_{\mr{travel},v}^{\mathrm{min}}$ & 2.5 & hours \\
      $\eta_{\mathrm{ch}}$ & 0.90 & --          & $\eta_{\mathrm{dis}}$ & 0.95 & -- \\
      \bottomrule
    \end{tabular}
  \end{adjustbox}
\end{table}
We used Yalmip~\cite{Lofberg2004} to formulate Problem 1 and employed Gurobi~\cite{gurobi} as the solver.
Solving the problem to a gap of less than 0.001\% took on average 14 seconds to parse and 1.3 minutes to solve on an Intel Core i7-12700H, 2.30 GHz processor with 16 GB RAM.

\section{Results}\label{sec:res}
In this section we provide our results for our four experiments. Experiment 1 corresponds to the expected design of the project. Experiment 2 shows the resulting design allowing for PV generation and an optimized peak power. Experiment 3 includes adds the possibility of a stationary battery. Experiment 4 additionally allows for optimized vessel battery design. The final designs and resulting costs are given in \ref{tab:system-config}.
Table~\ref{tab:system-config} summarizes the resulting designs and costs.
Then we describe the vessel results in Fig. \ref{fig:powers}, and then the shore-side results in Fig.~\ref{fig:port} under the cost optimal solution (Experiment 4).
\subsection{Designs and Costs}
\begin{table*}[!htbp]
  \centering
  \caption{System configuration for each experiment.}
  \label{tab:system-config}
  \renewcommand{\arraystretch}{1.1}
  \begin{adjustbox}{max width=\linewidth}
  \begin{tabular}{c c c c c c c c c c c}
  \toprule
  \textbf{Exp.} &
  \shortstack{\textbf{\(P^{\mathrm{M}}_{\mathrm{pv}}\)}\\\textbf{(MW)}} &
  \shortstack{\textbf{\(E_{i}^{\mathrm{M}}\)}\\\textbf{(MWh)}} &
  \shortstack{\textbf{Optimization of \(E_{v}\)}\\\textbf{(Boolean)}} &
  \shortstack{\textbf{Infrastructure}\\\textbf{Charging Power (MW)}} &
  \shortstack{\textbf{\(E_{v}\)}\\\textbf{(MWh)}} &
  \shortstack{\textbf{\(E_{\mathrm{b},i}\)}\\\textbf{(MWh)}} &
  \shortstack{\textbf{\(P_{\mathrm{pv},i}\)}\\\textbf{(MW)}} &
  \shortstack{\textbf{\(\mathbf{C^{-}}_{\mathrm{energy}}\)}\\\textbf{(kUSD)}} &
  \shortstack{\textbf{\(\mathbf{C^{+}}_{\mathrm{energy}}\)}\\\textbf{(kUSD)}} &
  \shortstack{\textbf{\(\mathbf{C}_{\mathrm{tot}}\)}\\\textbf{(kUSD)}} \\
  \midrule
  1 &
  0 &
  0 &
  0 &
  \{15, 15\} &
  \{40, 40\} &
  \{0, 0\} &
  \{0, 0\} &
  0&
  116.37&
  184.37\\
  2 &
  2.5 &
  0 &
  0 &
  \{8.72, 7.40\} &
  \{40, 40\} &
  \{0, 0\} &
  \{2.5, 2.5\} &
  4.56 &
  106.15 &
  169.65 \\
  3 &
  2.5 &
  50 &
  0 &
  \{8.04, 7.94\} &
  \{40, 40\} &
  \{3.51, 28.11\} &
  \{2.5, 2.5\} &
  0.74 &
  91.07 &
  165.78 \\
  4 &
  2.5 &
  50 &
  1 &
  \{8.04, 8.09\} &
  \{40.1, 35\} &
  \{3.51, 34.06\} &
  \{2.5, 2.5\} &
  0.88 &
  90.97 &
  164.78 \\
  \bottomrule
  \end{tabular}
  \end{adjustbox}
  \vspace{0.5em}
  \parbox{\textwidth}{
  \footnotesize
  \textit{Notes:} Experiment 1 shows the expected design of the project. Experiment 2 shows the design with only PV generation. Experiment 3 includes a PV system and a stationary battery. Experiment 4 includes a PV system, a stationary battery, and an optimized vessel battery.
}
\end{table*}
Experiment 1 represents the current design, without local EMS or PV generation. This configuration incurs the highest energy and infrastructure costs and an overall cost of \$184,370.
In Experiment 2, a maximum 2.5\unit{MW} PV system is introduced at each port. This addition results in a total cost reduction of \$14,720, driven by reduced energy costs. This is due to the reduced energy purchased () from the grid and extra revenue produced (). In all the remaining cases the final PV design was the upper limit $P^{\mathrm{M}}_{\mathrm{pv}}$.
Experiment 3 allows up to a maximum 50,000~kWh static battery to be used in the design. The final result are a battery of 3.51\unit{MWh} at BA, and a 28.1\unit{MWh} battery at CO. As a result, total weekly costs decrease by \$3,870 compared to experiment 2.
Finally, Experiment 4 builds upon the previous configuration by also optimizing vessel battery design. Despite a slight increase in the upfront battery cost, this configuration delivers the lowest total cost at \$164,780, achieving extra gains of \$1,000 compaerd to experiment 3.
In all cases the operational speed and travel time remain constant at 21.33 knots and 1.5 hours, respectively.
\subsection{Vessel Results}
\begin{figure}[!t]
  \centering
  \resizebox{\linewidth}{!}{%
      \includegraphics{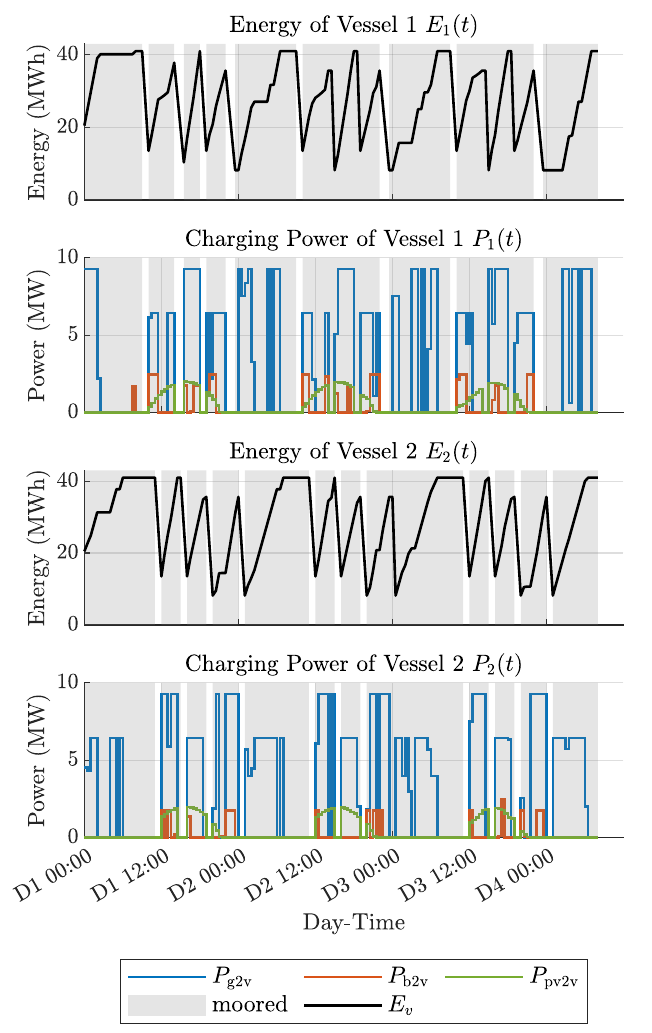}
  }
  \caption{Energy and power profiles for vessels \( V = 1 \) and \( V = 2 \) over a four-day horizon for experiment 4. The figure shows the battery energy \( E_v \) and the power flows \( P_{\mathrm{g2v}}, P_{\mathrm{b2v}}, \text{ and } P_{\mathrm{pv2v}}\), along with mooring periods.}
  \label{fig:powers}
\end{figure}

Figure~\ref{fig:powers} presents the operation curves for experiment 4. The figure illustrates the energy state \( E_v \) and power exchanges for each vessel. Grey regions indicate mooring periods during which vessels can charge.
The energy \( E_v \) exhibits a cyclic pattern aligned with mooring periods, with frequent charging events that bring the state-of-charge close to its upper bound.
The dominant power source is the grid (\( P_{\mathrm{g2v}} \)), whereas the battery-to-vessel (\( P_{\mathrm{b2v}} \)) and PV-to-vessel (\( P_{\mathrm{pv2v}} \)) flows are comparatively smaller and used during sunlight hours.
\subsection{Port Side Results}
\begin{figure}[!h]
  \centering
  \resizebox{\linewidth}{!}{%
      \includegraphics{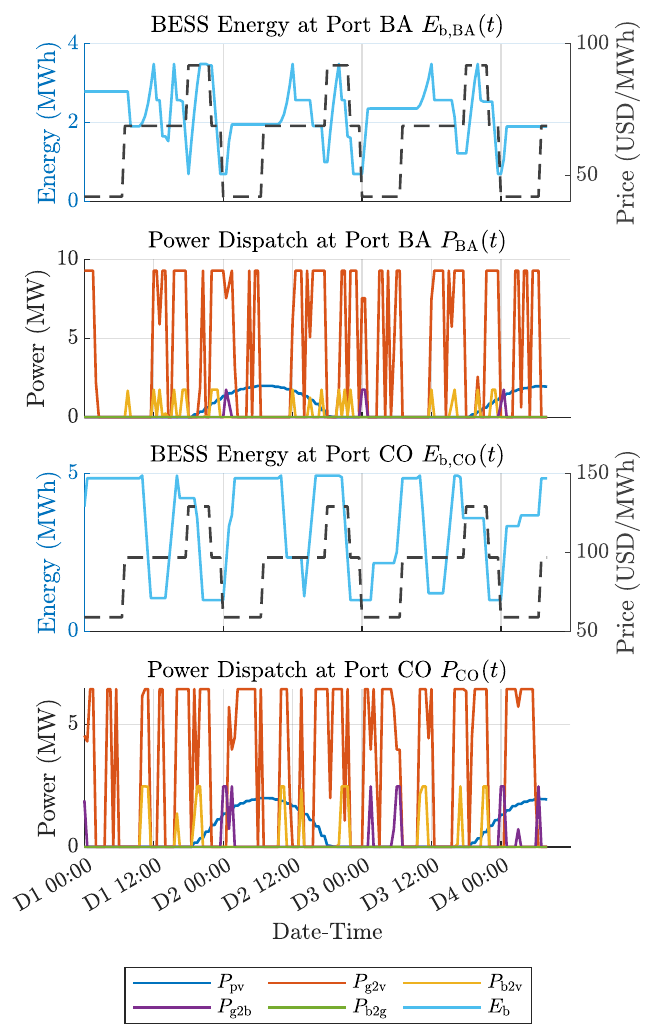}
  }
  \caption{Power flows and battery energy \( E_b \) at ports BA and CO over a four-day period for experiment 4. The power profiles include photovoltaic generation \( P_{\mathrm{pv}} \), grid-to-battery \( P_{\mathrm{g2b}} \), battery-to-grid \( P_{\mathrm{b2g}} \), grid-to-vessel \( P_{\mathrm{g2b}} \), and battery-to-vessel \( P_{\mathrm{b2v}} \).}
  \label{fig:port}
\end{figure}

Figure~\ref{fig:port} presents the energy and power dynamics at both ports.
We observe the grid-to-vessel power \( P_{\mathrm{g2v}} \) dominates, indicating a high dependence on the grid for direct vessel charging at both ports.
Nevertheless, the battery-to-vessel power \( P_{\mathrm{b2v}} \) is also utilized frequently, suggesting the battery is actively supporting charging.
Photovoltaic generation \( P_{\mathrm{pv}} \) is limited and concentrated during daytime, contributing marginally to the overall power demand.
Grid-to-battery \( P_{\mathrm{g2b}} \) and battery-to-grid \( P_{\mathrm{b2g}} \) flows are present but less frequent, suggesting the battery is primarily used for local vessel support rather than energy arbitrage.

\section{Conclusion}\label{sec:dis}
In this paper, we present an optimization framework aimed at improving the economic viability of electric-powered ship services by jointly optimizing charging schedules, infrastructure investments, and speed profiles.
Compared to an unoptimized scenario---with fixed maximum power station capacity and predetermined schedules based on planned itineraries---our method reduces total costs (comprising both capital and operational expenditures) by 7.8\%.
The proposed optimization framework not only serves as a design planning tool but also as an operational platform for weekly scheduling.
Future work will extend this study by incorporating power consumption variability along the route, moving beyond the current simplified modeling approach.
Additionally, we plan to integrate stochastic forecasts for weather and prices, as well as a more detailed treatment of battery degradation.
All results are heavily dependent on the prices used, thus a sensitivity analysis on the impact of all prices considered is warranted.
Overall, our results with conservative price values confirm that advanced optimization techniques can significantly improve both the upfront and total cost metrics, paving the way for more economically viable electric ship services.

\addtolength{\textheight}{0cm}%

\section*{Acknowledgment}

We thank Dr.~I.~New and F.~F.~Vehlhaber for proofreading this paper and providing valuable suggestions. This publication is part of the project GTD-Elektrifikatie, made possible in part by the Ministry of Economic Affairs and Climate Policy of the Netherlands.


\end{document}